\documentclass[notitlepage,twocolumn,pre,tightenlines,footinbib,superscriptaddress]{revtex4-1}

\usepackage{url}
\usepackage{gensymb}
\usepackage{amsmath}
\usepackage{amssymb,amsfonts}
\usepackage{graphicx}
\usepackage{subfigure}
\usepackage{color}

\newcommand{\mathnotation}[2]{\newcommand{#1}{\ensuremath{#2}}}
\mathnotation{\Sil}{\text{SiO}_2}
\mathnotation{\NC}{\text{Na}_2\text{CO}_3}
\mathnotation{\CC}{\text{CaCO}_3}

\begin{document}

\title{In-situ synchrotron microtomography reveals multiple reaction
pathways during soda-lime glass synthesis}

\author{E. Gouillart}
\affiliation{Surface du Verre et Interfaces, UMR 125 CNRS/Saint-Gobain,
  93303 Aubervilliers, France}
\author{M. J. Toplis}
\affiliation{
IRAP (UMR 5277, CNRS/University of Toulouse III), Observatoire Midi
Pyr\'en\'es, 14, Ave. E. Belin, 31400, Toulouse, France.
}
\author{J. Grynberg}
\author{M.-H. Chopinet}
\author{E. Sondergard}
\affiliation{Surface du Verre et Interfaces, UMR 125 CNRS / Saint-Gobain,
  93303 Aubervilliers, France}
\author{L. Salvo}
\author{M. Su\'ery}
\affiliation{SIMaP, UMR CNRS 5266, Grenoble INP, UJF, GPM2, BP 46, 38402
Saint-Martin d'H\`eres Cedex, France}
\author{M. Di Michiel}
\affiliation{ESRF, 156 rue des Martyrs, BP 220, 38043 Grenoble Cedex 9,
France}
\author{G. Varoquaux}
\affiliation{PARIETAL  (INRIA Saclay - Ile de France), NeuroSpin
CEA Saclay, Bat 145, 91191 Gif-sur-Yvette France}

\begin{abstract}

\bf

Ultrafast synchrotron microtomography has been used to study
\emph{in-situ} and in real time the initial stages of silicate glass melt
formation from crystalline granular raw materials. Significant and
unexpected rearrangements of grains occur below the nominal eutectic
temperature, and several drastically different solid-state reactions are
observed to take place at different types of intergranular contacts.
These reactions have a profound influence on the formation and the
composition of the liquids produced, and control the formation of
defects.

\end{abstract}

\maketitle 

\begin{figure}
\centerline{
    \includegraphics[width=0.99\columnwidth]{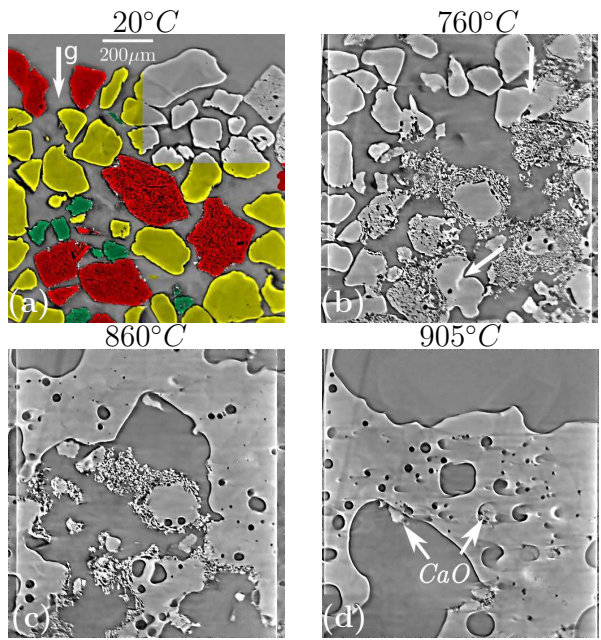}
}
\caption{ (a) Slice through the reconstructed 3-D volume of the pile of
raw materials at room temperature. The segmented volumes of sand, sodium
carbonate and calcium carbonate grains have been coloured in yellow, red
and green respectively. The top-right quarter has been left uncolored to
show the original slice. (b-d) Slice through the same plane as in (a), at
$760\,\degree$, $860\,\degree$, and $905\,\degree$C. Fragmented sodium
carbonate grains react with sand to produce porous crystalline silicates
(panel b), and rare liquid bridges (white arrow
in panel b). Further increase in temperature leads to formation of a
significant proportion of liquid (panel c), finally reaching a state in
which the the granular packing is transformed into a viscous melt
with bubbles and grain inclusions (d). \label{fig:system}}

\end{figure}

\section{Introduction}

Many commonplace materials are manufactured from a loose packing of
coarse reactive grains. Among such materials, window-glass production
relies on synthesis from a mixture of quartz sand, sodium carbonate, and
calcium carbonate.  Industrial synthesis of good-quality soda-lime glass
is generally carried out at $1400-1500\,\degree$C, despite the fact that
typical compositions are completely molten at $1050\,\degree$C. Such high
temperatures are required to eliminate defects (gas bubbles and unmolten
sand grains), and to homogenize the melt. Parameters such as the grain
size of raw materials are known to influence the quality of
glasses~\cite{Potts1944, Tooley1946, Wilburn1965, Sheckler1990}. However,
little direct information is available concerning the complex interplay
between the geometry of the system and the rate of chemical reactions
taking place during the earliest stages of melting. 

In this respect, \emph{in-situ} high-temperature tomographic X-ray
imaging~\cite{Baruchel2006} is a potentially
powerful technique as it provides the possibility to: a) quantitatively
describe the distribution of grains and the nature of solid-solid
contacts in the initial pile; b) identify where and at what temperature
reactions between grains occur; c) quantify the extent to which these
reactions occur. This technique has been employed in the past to study
the evolution of powder compacts during the sintering of glass
beads~\cite{Bernard2005} or metallic powders~\cite{Lame2004}. However, no
tomographic data have been acquired concerning a high temperature system
involving chemical reactions between grains and the irreversible
formation of a liquid. 

\section{Experimental procedure}

An \emph{in-situ} microtomography experiment has been performed on the
ID15A beamline at the European Synchotron Radiation Facility (ESRF). We
used white X-ray radiation with a peak photon energy of 40 keV. The
spatial resolution was $1.6 \mu\text{m}$. A mixture of Ronceveaux
silica sand, Solvay$\circledR$ sodium carbonate and Saint-Germain calcium
carbonate was poured into a 2-mm-diameter silica crucible. Weight
percentages of these three materials were 64, 19 and 17 respectively, and
each material had a characteristic grain-size of: $160-200\,\mu\text{m}$,
$250-320\,\mu\text{m}$, and $80-100\,\mu\text{m}$ for sand, sodium
carbonate and calcium carbonate respectively (see Fig.~\ref{fig:system}a
and Supplementary Movie). Such grain sizes are consistent with those used
by the glass making industry. The sample was first heated from room
temperature to $740\,\degree$C at a rate of 70 $\degree$C/min, during
which rapid thermal dilation prevented imaging. Above
that temperature, the sample was heated at 5 $\degree$C/min up to
$1100\,\degree$C. 3-D scans of the sample were acquired every
$2.6\,\degree$C, providing a direct picture of the evolution of the
microstructure of the system, as illustrated in
Fig.~\ref{fig:system}(a-d) (see also Supplementary Movie). 

In order to obtain quantitative information, 3-D volumes have been
processed and segmented (see Supplementary Materials). All of the 175
grains in the analyzed volume at room temperature have been identified
(see Fig.~\ref{fig:system}a). Sodium carbonate grains are characterized
by their large internal porosity (red grains in
Fig.~\ref{fig:system}a).

\section{Results and discussion}
 
\begin{figure}
\centerline{\includegraphics[width=0.99\columnwidth]{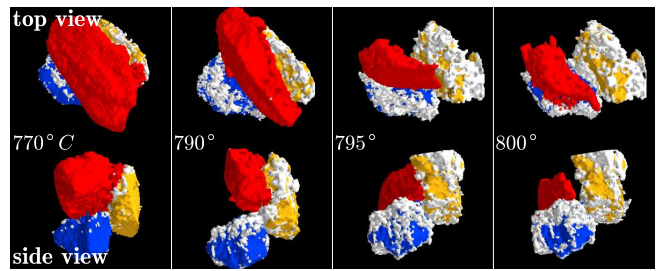}}
\caption{Top- and side-view of one sodium carbonate grain (red) and two
sand grains (yellow and blue) at four different temperatures, that react
together to form silicates (white). For the sake of clarity, neighboring
grains have not been represented. \label{fig:scenar_carbo}}
\end{figure}

\begin{figure}
\centerline{
    \includegraphics[width=0.95\columnwidth]{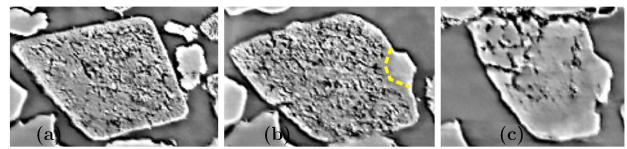}
}

\caption{(a-c) Evolution of two neighboring grains of sodium carbonate
(large grain at the center) and calcium carbonate (small grain on the
right) at room temperature, $760\,\degree$C, and $825\,\degree$C. The
calcium carbonate grain (delineated in yellow in (b) and indicated by the
presence of additional facets) is incorporated into its larger neighbor,
suggesting the formation of a double carbonate. At $825\,\degree$C, the
eutectic melting of sodium carbonate and the double carbonate produces
a carbonate liquid, as shown by the absence of porosity on the right part
of the grain. \label{fig:calcium}}

\end{figure}

Our data reveal the importance of solid-state reactions before the first
appearance of melts. Between room temperature and $750\,\degree$C, many
sodium carbonate grains have unexpectedly broken up
(Fig.~\ref{fig:system}b). This dramatic change in morphology continues
with increasing temperature, as illustrated in
Fig.~\ref{fig:scenar_carbo} for the reaction between a sodium carbonate
grain (in red), and two sand grains (in yellow and blue). As
temperature increases, the original sodium carbonate grain shrinks, with
the appearance of a jagged and fuzzy reaction layer (shown in white).
This newly formed material sticks to either of the adjacent sand grains
and the residual carbonate grain moves, seemingly attracted to
unreacted parts of the sand grains. A large fraction of the sand grains
is rapidly covered in a loose shell of the new high porosity material
(Fig.~\ref{fig:scenar_carbo}). Consideration of the relevant phase
diagrams (see Supplementary Materials) and data from the literature
including thermogravimetric experiments~\cite{Wilburn1965}, in-situ X-ray
diffraction~\cite{Dolan2004} and in-situ NMR~\cite{Jones2005} all suggest
that formation of crystalline sodium silicates
($\text{Na}_2\text{SiO}_3$ and/or $\text{Na}_2\text{Si}_2\text{O}_5$) is
responsible for this fragmentation. Furthermore, a second type of
solid-state reaction occurs involving the two carbonates. Although only a
minority fraction is concerned, some grains of $\CC$ grains are found to
react with $\NC$ at temperatures below $750\,\degree$C to form a new
compound (Fig.~\ref{fig:calcium}a, b). It has been suggested in the
literature that formation of a Na-Ca double carbonate is
possible~\cite{Chopinet2010}, but no direct evidence for such a reaction
had been found when silica is present.

\begin{figure}
\centerline{
    \includegraphics[width=0.99\columnwidth]{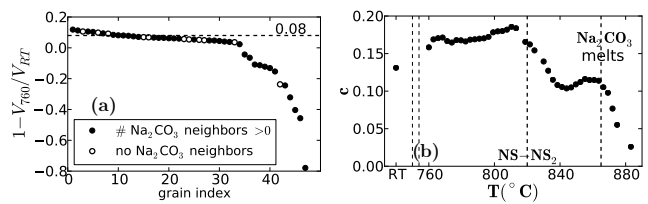}
}
\caption{(a) Relative volume growth of sand grains between room
temperature and $760\,\degree$C, sorted by increasing volume loss. Full
(resp. hollow) circles correspond to sand grains with (resp. without)
sodium carbonate neighbors. 
(b) Evolution of the fraction
$c$ of solid sodium carbonate and silicate phase with increasing
temperature. Two major drops occur around the eutectic temperatures
between sodium metasilicate (NS) and sodium disilicate ($\text{NS}_2$),
and the melting temperature of sodium carbonate. 
\label{fig:sand_attacks}}
\end{figure}

Inspection of the different evolutions of grains shows that the
occurrence of such solid-state reactions is conditioned by the proximity
of sand and calcium carbonate to sodium carbonate. Consideration of the
statistics of contact types demonstrates that in the initial packing,
28 sand grains (out of 108) and 18 calcium carbonate grains (out of 56)
did not have any contact with a sodium carbonate grain. The volumes of 47
individual silica grains have been compared at room temperature and at
$760\,\degree$C (Fig.~\ref{fig:sand_attacks}a). These grains are a
mixture of grains with and without contacts with sodium carbonate. We
find that a little over 30 of these grains show a volume increase of $8\%
\pm 2 \%$ (dotted line in Fig.~\ref{fig:sand_attacks}a) corresponding to
that expected from the $\alpha$ to $\beta$ transition of quartz. On the
other hand, approximately one third of the original sand grains have
experienced a volume increase smaller than 8\%, or even a significant
volume decrease. The grains concerned are found to be systematically
surrounded by the fine grained phase produced from sodium carbonate. It
is found that at least one sodium carbonate neighbor at room temperature
is a necessary but not sufficient condition for a sand grain to have
decreased in size at $760\,\degree$C. Only about one half of the original
sand-sodium carbonate contacts are concerned by a loss in volume of the
sand grain at $760\,\degree$C, probably because some of the contacts are
lost when sodium carbonate starts reacting and moving, as illustrated in
Fig.~\ref{fig:scenar_carbo}. Concerning calcium carbonate, given the
large fraction of sand and the reaction-induced elimination of sodium
carbonate (Fig.~\ref{fig:scenar_carbo}), the majority of calcium
carbonate grains are not in contact with sodium carbonate. Such grains
isolated from $\NC$ are observed to simply lose their $\text{CO}_2$,
producing refractory grains of $\text{CaO}$ (Fig.~\ref{fig:system}d), in
contrast to the reactive pathway shown in Fig.~\ref{fig:calcium}. 

The appearance of the first melts appears to be determined by the
solid-state reactions. Even at $760\,\degree$C certain sand grains show
textural evidence for the presence of a small amount of liquid
(highlighted by the arrows in Fig. ~\ref{fig:system}b), despite the
fact that the lowest stable eutectic in the
$\text{Na}_2\text{O}-\text{SiO}_2$ system is at
$790\,\degree$C~\cite{Kracek1930, Zaitsev1999}. The presence of liquid at
such low temperature may be explained by the existence of a metastable
eutectic between sodium metasilicate and silica (see Supplementary
materials). Quantification of the amount of liquid is not possible
directly, but may be estimated indirectly from the evolution of the
amount of texturally distinctive sodium carbonate and associated
secondary phases (e.g. Fig.~\ref{fig:system}b). The proportion of these
porous phases increases up to $820\,\degree$C
(Fig.~\ref{fig:sand_attacks}b), interpreted to reflect the conversion of
sand grains to sodium silicates. From $760$ to $820\,\degree$ C, we
observe in places a scarce production of liquid between some of the
sand grains; nevertheless, the production of crystalline phases is much
more important than the production of liquid, as shown by
Fig.~\ref{fig:sand_attacks}b. Despite the large fraction of silica in the
final glass composition (75\%), we do not observe any significant
formation of melts around $790\,\degree$C, the eutectic temperature
between silica and sodium disilicate. However, at $820\,\degree$C there
is a sharp drop in the proportion of porous phases, and clear
appearance of a significant amount of liquid around sand grains
previously covered by a layer of fine-grained porous phase
(Fig.~\ref{fig:system}b). This temperature corresponds to the equilibrium
eutectic between sodium metasilicate and sodium disilicate. The formation
of this sodium-rich liquid is less favorable for glass melting than
formation of the silica-sodium disilicate eutectic, which contains more
silica, thus resulting in greater consumption of crystalline silica.
These observations therefore provide strong indirect evidence in favor of
the low temperature formation of both crystalline
$\text{Na}_2\text{SiO}_3$ and $\text{Na}_2\text{Si}_2\text{O}_5$. An even
more abrupt decrease in the fraction of the fine grained porous phase is
observed at $865\,\degree$C (Fig.~\ref{fig:sand_attacks}b), corresponding
to the melting of sodium carbonate. The proportion of porous solids drops
rapidly to zero afterwards. Also, visual inspection reveals that at
$825\,\degree$C, the porosity of the few sodium carbonate grains that
have cemented to neighboring calcium carbonate grains
(Fig.~\ref{fig:calcium}b) is suddenly invaded by the presence of a liquid
(Fig.~\ref{fig:calcium}c). This temperature agrees well with the
incongruent melting temperature of the Na-Ca double carbonate at
~$820\,\degree$C~\cite{Chopinet2010, Gadalla1984}. 
 
Finally, our data offer insights into the generation of defects. Quartz
grains not surrounded by porous sodium carbonate or silicates below
$820\,\degree$C are found to have approximately the same size at
$950\,\degree$C as at room temperature. It is undoubtedly this population
of grains that will remain as high temperature solid defects because of
slow local diffusion. In the same way, enhancing the formation of Na-Ca
double carbonate will act to eliminate the generation of CaO.

In summary, these \emph{in-situ} observations combined with quantitative
image processing reveal unprecedented processes of glass melting. From a
physical point of view, the importance of direct observation is
exemplified by the unexpected and dramatic effect of solid-state
reactions on the spatial distribution of sodium carbonate, changes in
microstructure that in turn lead to accelerated reactions compared to a
fixed geometry. From a chemical point of view, many excursions from
overall thermodynamic equilibrium are observed. Short range packing
arrangements have a profound influence on local reaction pathways, as
most eloquently illustrated by the divergent fates of different calcium
carbonate grains. The occurence of different reactions at different
places strongly encourages the use of a spatially-resolved technique such
as tomographic imaging in order to study glass melting. Other
metastable phenomena include the generation of liquids well below those
predicted by the equilibrium phase diagram, and the observation of a
eutectic transition in the $\text{Na}_2\text{O}$-$\text{SiO}_2$ phase
diagram, which is not the closest eutectic to the mean composition.

\end{document}